\documentclass[aps,preprintnumbers,showpacs,twocolumn,superscriptaddress,floatfix,nofootinbib]{revtex4-1}

\usepackage{latexsym} 
\usepackage{amssymb} 
\usepackage{amsfonts}
\usepackage{amsmath} 
\usepackage{bm} 
\usepackage{mathtools}
\usepackage{envmath}
\usepackage{color}
\usepackage{times} 
\usepackage{hyperref}
\usepackage{float}
\usepackage{latexsym}
\usepackage{envmath}

\def\prl{Phys. Rev. Lett.}
\def\prd{Phys. Rev. D}
\def\cqg{Class. Quantum Grav.}

\def\apj{Astrophys. J.}

\usepackage{amssymb,amsmath}
\usepackage{graphicx} 
\begin{document}
\title{Critical phenomena in the aspherical gravitational collapse of radiation fluids}

\author{Thomas W. \surname{Baumgarte}}\affiliation{Department of Physics and Astronomy, Bowdoin College, Brunswick, ME 04011, USA}

\author{Pedro J. \surname{Montero}}\affiliation{Max-Planck-Institute f\"ur Astrophysik, Karl-Schwarzschild-Str.~1, 85748 Garching bei M\"unchen, Germany}

%%%%%%%%%%%%%%%
%%% Abstract
%%%%%%%%%%%%%%%

\begin{abstract}
We study critical phenomena in the gravitational collapse of a radiation fluid.  We perform numerical simulations in both spherical symmetry and axisymmetry, and observe critical scaling in both supercritical evolutions, which lead to the formation of a black hole, and subcritical evolutions, in which case the fluid disperses to infinity and leaves behind flat space.   We identify the critical solution in spherically symmetric collapse, find evidence for its universality, and study the approach to this critical solution in the absence of spherical symmetry.   For the cases that we consider, aspherical deviations from the spherically symmetric critical solution decay in damped oscillations in a manner that is consistent with the behavior found by Gundlach in perturbative calculations.  Our simulations are performed with an unconstrained evolution code, implemented in spherical polar coordinates, and adopting ``moving-puncture" coordinates.
\end{abstract}


\maketitle

%%%%%%%%%%%%%%%%%%%%%%%%%%%%%%%%%%%%%%%%%%%%%%%%%%%%%%%%%%%%
\section{Introduction}
\label{sec:intro}
%%%%%%%%%%%%%%%%%%%%%%%%%%%%%%%%%%%%%%%%%%%%%%%%%%%%%%%%%%%%

Critical phenomena in gravitational collapse were first reported in the seminal work of Chopuik \cite{Cho93}.   Shortly after this original discovery, which was based on studies of massless scalar fields in spherical symmetry, similar behavior was found in other matter models, including vacuum (i.e.~pure gravitational waves) \cite{AbrE93}, and radiation fluids \cite{EvaC94} (see the excellent reviews \cite{Gun03,GunM07} for a much more comprehensive discussion.)

Critical collapse can be observed in the evolution of generic initial data close to the threshold of black-hole formation.   We again refer to \cite{Gun03,GunM07} for thorough reviews, and briefly summarize only the most important characteristics here.  Consider initial data that are parametrized by some parameter $p$, with the evolution of data with $p$ greater than some critical parameter $p_\star$ leading to black-hole formation.   Close to the critical parameter the following critical phenomena can then be observed in the resulting spacetimes.   For $p > p_\star$, the mass of the newly formed black holes scales with
\begin{equation} \label{scaling}
M \propto (p - p_\star)^\gamma,
\end{equation}
where the critical exponent $\gamma$ depends on the matter model, but not on the specific choice or parametrization of the initial data.   In the strong-field region prior to black-hole formation the spacetime approaches a self-similar critical solution, which again depends on the matter model only.  For some matter models, including massless scalar fields, the critical solution features a discrete self-similarity, while others, including perfect fluids, feature a continuous self-similarity.

Numerous studies, both numerical and analytical, have followed up on the above initial reports of critical phenomena.  Most of this work, however, has focused on spherical symmetry, where the demand for high spatial resolution of increasingly small features can be met most easily.  As a consequence, aspects of critical collapse that can be studied only in the absence of spherical symmetry remain largely unexplored, for example the effects of angular momentum (but see \cite{ChoHLP04}, as well as \cite{GunM07} for a summary of results from perturbative calculations.)  Another important question remains unresolved: studying linear perturbations of the spherically symmetric critical solution for scalar fields, Mart\'in-Garc\'ia and Gundlach \cite{MarG99} found that all non-spherical perturbations decay, while Choptuik {\it et.al.}~\cite{ChoHLP03b} found numerical evidence for the existence of an aspherical growing mode.   Given the richness of the subject, the number of numerical studies of critical collapse in the absence of spherical symmetry is surprisingly small (some recent examples of numerical studies of critical collapse in the absence of spherical symmetry include \cite{Sor11,HilBWDBMM13,HeaL14}.)  As Gundlach and Mart\'in-Garc\'ia observed in their review \cite{GunM07}, ``there has been less progress in going beyond spherical symmetry than we anticipated."  

In the meantime, numerical relativity simulations in three spatial dimensions have made tremendous progress following the first successful simulations of the inspiral of binary black holes \cite{Pre05a,CamLMZ06,BakCCKM06a}.  While these newly available codes have been used to study many interesting astrophysical processes, they have rarely been applied to studies of critical collapse (see \cite{HilBWDBMM13,HeaL14} for two examples.)  One possible reason is that most of these codes are based either on Cartesian coordinates or more complicated coordinate patches, neither one of which are well suited for simulations of the small spatial structures encountered in critical collapse simulations.   Presumably, spherical polar coordinates are better suited to study critical phenomena, in particular the behavior of deviations from the spherically symmetric critical solution (see, e.g., the discussion in \cite{ChoHLP03b}.)   We have recently developed a numerical code that solves Einstein's equations in spherical polar coordinates \cite{MonC12,BauMCM13,BauMM15}, and in this paper we use this code to study the critical collapse of radiation fluids.

The original discovery of critical phenomena in the collapse of a radiation fluid by Evans and Coleman \cite{EvaC94} was quickly followed by a number of both analytical (e.g.~\cite{KoiHA95,Mai96}) and numerical studies (e.g.~\cite{NeiC00b}.)  To the best of our knowledge, however, all of these studies were performed under the assumption of spherical symmetry.  Here we generalize some of these results by relaxing this assumption; in particular we compute scaling laws of the form (\ref{scaling}) for axisymmetric data.  As first discussed in \cite{EvaC94}, the critical solution encountered in the collapse of radiation fluids features a continuous self-similarity (CSS), rather than the discrete self-similarity found in many other matter models.   We identify this critical solution in our spherically symmetric numerical solutions, and study the approach to this critical solution in our aspherical simulations.  Close to criticality, the aspherical solutions perform a damped oscillation around the spherical critical solution in a manner very similar to that described by Gundlach \cite{Gun98b,Gun02} -- at least for the cases that we considered.  We believe that this is the first confirmation of this behavior in a non-linear numerical simulation.

In addition to analyzing critical phenomena in the collapse of a radiation fluid in the absence of spherical symmetry, this paper serves to test and calibrate the performance of a free (i.e.~unconstrained) evolution code in the context of critical collapse.  Traditionally, most codes used in simulations of critical collapse were specifically designed for that purpose: they made symmetry assumptions, adopted specific slicing conditions (e.g.~maximal or polar slicing) and used constrained evolution, in which Einstein's constraint equations are used to replace at least some of the evolution equations.  Many more recently developed codes used in simulations of binary coalescence, on the other hand, are designed very differently: they do not make any symmetry assumptions and use free evolution, in which the constraint equations can be monitored but are not solved.  In general these codes can be run with different slicing and gauge conditions, but the  ``1+log" and ``Gamma-driver" conditions (see eqs.~(\ref{1+log}) and (\ref{Gamma-driver}) below) have proven particularly useful for simulations of spacetimes containing black holes.   How suitable these codes are for simulations of critical collapse, however, remains a somewhat open question (see also \cite{HilBWDBMM13,HilWB15,AkbC15} for recent discussions.)  Our findings here demonstrate that, at least for some matter models, unconstrained evolution codes with the 1+log and Gamma-driver coordinate conditions can indeed be used to study critical phenomena.

Our paper is organized as follows.  In Section \ref{sec:setup} we describe the setup of the problem, including a brief description of our numerical methods and the form of our initial data.  We discuss our numerical results in Section \ref{sec:results}.  In Section \ref{sec:ss_centered} we focus on ``centered" initial data, which take their maximum density at the origin of the coordinate system.  These simulations allow us to compare directly with the findings of \cite{EvaC94}.  In order to study deviations from spherical symmetry, however, it also makes sense to study ``off-centered" initial data, which take their maximum away from the origin.  We first consider such data in spherical symmetry in Section \ref{sec:ss_offcenter}, and then generalize these to axisymmetry in Section \ref{sec:axi}.  We conclude with a brief discussion in Section \ref{sec:discussion}.  We also include an appendix with some details on the logarithmic radial grid used in this paper.

%%%%%%%%%%%%%%%%%%%%%%%%%%%%%%%%%%%%%%%%%%%%%%%%%%%%%%%%%%%%
\section{Setup of the Problem}
\label{sec:setup}
%%%%%%%%%%%%%%%%%%%%%%%%%%%%%%%%%%%%%%%%%%%%%%%%%%%%%%%%%%%%

%%%%%%%%%%%%%%%%%%%%%%%%%%%%%%%%%%%%%%%%%%%%%%%%%%%%%%%%%%%%
\subsection{Basic Equations and Numerical Solution}
\label{subsec:basics}
%%%%%%%%%%%%%%%%%%%%%%%%%%%%%%%%%%%%%%%%%%%%%%%%%%%%%%%%%%%%

In the following we construct numerical solutions of Einstein's equations
\begin{equation} \label{einstein}
G_{ab} = 8 \pi T_{ab}
\end{equation}
for the stress-energy tensor
\begin{equation} \label{stress_energy}
T_{ab} = (\rho + P) u_a u_b + P g_{ab}
\end{equation}
describing a perfect fluid, where we have adopted geometrized units with $G = c = 1$.   Here $G_{ab}$ is the Einstein tensor, $\rho$ is the total energy density, $P = (\gamma - 1) \rho$ is the pressure, $u^a$ is the fluid four-velocity, and $g_{ab}$ is the spacetime metric.   We specialize to a radiation fluid, for which $\gamma = 4/3$.  

We solve Einstein's equations (\ref{einstein}) using the Baumgarte-Shapiro-Shibata-Nakamura (BSSN) formulation \cite{NakOK87,ShiN95,BauS98}.  The BSSN formulation employs a 3+1 decomposition of the spacetime, by which the spacetime metric $g_{ab}$ induces a spatial metric 
\begin{equation}
\gamma_{ab} = g_{ab} + n_a n_b
\end{equation}
on the spatial slices (see, e.g., \cite{BauS10} for a textbook treatment.)  Here $n^a$ is the normal on the spatial slice, which, in terms of the lapse function $\alpha$ and the shift vector $\beta^i$, may be written as
\begin{equation}
n^a = \alpha^{-1} (1, -\beta^i)
\end{equation}
(here and in the following indices $a, b, c, \ldots$ denote spacetime indices, while $i, j, k, \ldots$ denote spatial indices.)   
% We denote the relativistic gamma-factor between a normal observer and an observer comoving with the fluid as
% \begin{equation} \label{W}
% W \equiv - n_a u^a = \alpha u^t.
% \end{equation} 
Another important quantity is the extrinsic curvature, which may be written as
\begin{equation}
K_{ij} = - \frac{1}{2 \alpha} \partial_t \gamma_{ij} + D_i \beta_j + D_j \beta_i,
\end{equation}
where $D_i$ denotes the covariant derivative associated with $\gamma_{ij}$, as well as its trace $K \equiv \gamma^{ij} K_{ij}$. 

The formalism also adopts a conformal rescaling of the spatial metric $\gamma_{ij}$,
\begin{equation}
\gamma_{ij} = \psi^4 \bar \gamma_{ij},
\end{equation}
where $\psi$ is the conformal factor and $\bar \gamma_{ij}$ the conformally related metric.  In the BSSN formalism the conformal factor is usually written as $\psi = e^\phi$.

In order to implement the BSSN formalism in spherical polar coordinates we also employ a reference-metric approach \cite{BonGGN04,ShiUF04,Bro09,Gou12}.   In such a reference-metric approach, some geometric objects associated with the conformally related metric $\bar \gamma_{ij}$ are expressed in terms of the difference between these objects and their counterparts associated with a reference metric $\hat \gamma_{ij}$ (see equation (\ref{lambda}) below for an example).  For our applications it is natural to choose $\hat \gamma_{ij}$ to be the flat metric expressed in spherical polar coordinates.  We further scale out appropriate powers of the geometric factors $r$ and $\sin \theta$ from all tensorial quantities, so that for regular spacetimes all dynamical variables used in the code remain regular.  We do not, however, attempt to regularize the equations, which still contain inverse powers of $r$ and $\sin \theta$ and hence become singular at the origin and on the axis.   We adopt a finite-differencing method and evolve the resulting equations with a partially implicit Runge-Kutta (PIRK) time integration method \cite{MonC12,CorCD14}.  Details of our numerical implementation can be found in \cite{BauMCM13,BauMM15}.   

For all simulations reported in this paper we impose ``moving puncture" coordinate conditions, i.e.~the 1+log condition for the lapse function $\alpha$
\begin{equation} \label{1+log}
( \partial_t - \beta^j \partial_j) \alpha = - 2 \alpha K
\end{equation}
(see \cite{BonMSS95}) together with a version of a Gamma-driver condition
\begin{equation} \label{Gamma-driver}
( \partial_t - \beta^j \partial_j) \beta^i = \mu_S \bar \Lambda^i
\end{equation}
with $\mu_S = 3/4$ for the shift vector $\beta^i$ (see \cite{AlcBDKPST03,ThiBB11}.)  Here the $\bar \Lambda^i$ play the role of the conformal connection functions,
\begin{equation} \label{lambda}
\bar \Lambda^i \equiv \bar \gamma^{jk} (\bar \Gamma^i_{jk} - \hat \Gamma^i_{jk}),
\end{equation}
where $\bar \Gamma^i_{jk}$ and $\hat \Gamma^i_{jk}$ are the connection symbols associated with the conformally related metric $\bar \gamma_{ij}$ and the flat reference metric $\hat \gamma_{ij}$, respectively.

We solve the fluid equations by applying a similar reference-metric approach to the ``Valencia"-form \cite{BanFIMM97} of the equations of relativistic hydrodynamics (see also \cite{MonBM14,BauMM15}.)   For the purposes in this paper the equations simplify, since, for a radiation fluid, the rest density $\rho_0$ vanishes, and the total energy density $\rho$ is given by the internal energy density alone.  We solve the resulting equations using a high-resolution shock-capturing technique as described in \cite{MonBM14,BauMM15}.

While our code was originally designed for three spatial dimensions without any symmetry assumptions, we here specialize to axisymmetric spacetimes, so that our solutions depend on a radius $r$ and an angle $\theta$ only.   We also impose equatorial symmetry, so that we can restrict our computations to one hemisphere.  One significant enhancement used in this paper concerns the radial differencing.  While we used a uniform grid in all coordinates in the above references, we here allow for a logarithmic grid in the radial directions.  We provide more details on both the grid setup and the finite differencing stencils in Appendix \ref{app:log}.  

For the simulations presented in this paper we impose outer boundary conditions at $r_{\rm out} = 32$ (in our code units; see below), and choose each radial grid cell to be larger than its smaller-radius neighbor by a factor of $c = 1.02$.  We also use two different radial grid resolutions.  In what we refer to as ``high-resolution" runs we use $N_r = 396$, which results in a grid spacing  of $\Delta_2 = 2.44 \times 10^{-4}$ across the origin (see Fig.~\ref{Fig:grid}) and a grid spacing at the outer boundary of 0.634.  To put this into perspective, an AMR application, in which each refinement level has twice the grid resolution of the previous level, would require about 12 nested grid levels to achieve a similar range in grid resolution.   In our "low-resolution" runs we use $N_r = 288$, for which  $\Delta_2 = 2.08 \times 10^{-3}$, and the grid spacing at the outer boundary is 0.636.  The structure of our gird is also visible in Figs.~\ref{Fig:rho_axi} and \ref{Fig:Omega_profiles} below.

%%%%%%%%%%%%%%%%%%%%%%%%%%%%%%%%%%%%%%%%%%%%%%%%%%%%%%%%%%%%
\subsection{Initial Data}
\label{sec:indata}
%%%%%%%%%%%%%%%%%%%%%%%%%%%%%%%%%%%%%%%%%%%%%%%%%%%%%%%%%%%%

We start our simulations with initial data that are both conformally flat, i.e.~$\bar \gamma_{ij} = \hat \gamma_{ij}$, and time-symmetric, i.e.~$K_{ij} = 0$.  We then specify the initial density distribution as follows,
\begin{equation} \label{rho_init}
\rho(r,\theta) = \frac{\eta}{4 \pi^{3/2} R_0^2} \left(1.0 + \epsilon \frac{r^2 } {R_0^2 + r^2}  P_2(\theta)\right) \left(f_+ + f_-\right)
\end{equation} 
%\begin{equation} \label{rho_init}
%\rho(r,\theta) = \eta \left(1.0 + \epsilon \frac{r^2}{R_0^2 + r^2} P_2(\theta) \right) 
%\end{equation} 
where we have abbreviated
\begin{equation}
f_\pm=  
\exp \left( - \frac{ (\psi^2 r \pm R_c)^2}{R_0^2} \right),
\end{equation}
and where 
\begin{equation}
P_2(\theta) = \frac{1}{2} \left( 3 \cos^2 \theta - 1 \right)
\end{equation}
is the second-order Legendre polynomial.  In the above expression, $\eta$ parameterizes the overall amplitude of the density, while $\epsilon$ determines deviations from spherical symmetry.  In $f_\pm$ we have multiplied the (isotropic) radius $r$ with the square of the conformal factor, $\psi^2$, so that, in spherical symmetry, the product can be identified with the areal radius $R$.  The density distribution is then centered on an areal radius of approximately $R_c$, while its length-scale is approximately $R_0$.  In the following we will consider different choices for $\eta$, $\epsilon$ and $R_c$, but will always use $R_0 = 1$, which determines our code units.  In the following, all dimensional quantities are expressed in units of $R_0$.

The density distribution (\ref{rho_init}) depends on the conformal factor, which is found by solving the Hamiltonian constraint
\begin{equation}\label{Hamiltonian}
\nabla^2 \psi = - 2 \pi \psi^5 \rho.
\end{equation}
In practice we iterate between the two equations (\ref{rho_init}) and (\ref{Hamiltonian}) until convergence has been achieved.

For spherically symmetric and centered data, i.e.~$\epsilon = R_c = 0$, the density distribution (\ref{rho_init}) reduces to 
\begin{equation} \label{evanscoleman}
\rho(r) = \frac{\eta}{2 \pi^{3/2} R_0^2} \exp(- R^2/R_0^2),
\end{equation}
which is the initial density distribution adopted in \cite{EvaC94} (see their equation (6).)  The total gravitational mass $M$ can then be computed analytically,
\begin{equation} \label{ADM_mass}
M = \frac{1}{2} \eta R_0~~~~~~~~~~~(\mbox{for}~\epsilon = R_c = 0),
\end{equation}
which shows that $\eta = 2 M / R_0$ serves as a non-dimensional measure of the strength of the gravitational fields.  In generalizing the density (\ref{evanscoleman}) to the aspherical distributions (\ref{rho_init}) we have introduced extra factors that ensure that the density is smooth at the origin.

Finally, we initialize the lapse function $\alpha$ and the shift vector $\beta^i$ according to
\begin{equation}
\alpha = \psi^{-2},~~~~~~~~\beta^i = 0.
\end{equation}

%%%%%%%%%%%%%%%%%%%%%%%%%%%%%%%%%%%%%%%%%%%%%%%%%%%%%%%%%%%%
\section{Results}
\label{sec:results}
%%%%%%%%%%%%%%%%%%%%%%%%%%%%%%%%%%%%%%%%%%%%%%%%%%%%%%%%%%%%

%%%%%%%%%%%%%%%%%%%%%%%%%%%%%%%%%%%%%%%%%%%%%%%%%%%%%%%%%%%%
\subsection{Spherical Symmetry: $\epsilon = 0$}
\label{sec:ss}
%%%%%%%%%%%%%%%%%%%%%%%%%%%%%%%%%%%%%%%%%%%%%%%%%%%%%%%%%%%%

%%%%%%%%%%%%%%%%%%%%%%%%%%%%%%%%%%%%%%%%%%%%%%%%%%%%%%%%%%%%
\subsubsection{Centered Data: $R_c = 0$}
\label{sec:ss_centered}
%%%%%%%%%%%%%%%%%%%%%%%%%%%%%%%%%%%%%%%%%%%%%%%%%%%%%%%%%%%

\begin{figure}
\begin{center}
\includegraphics[width=3.5in]{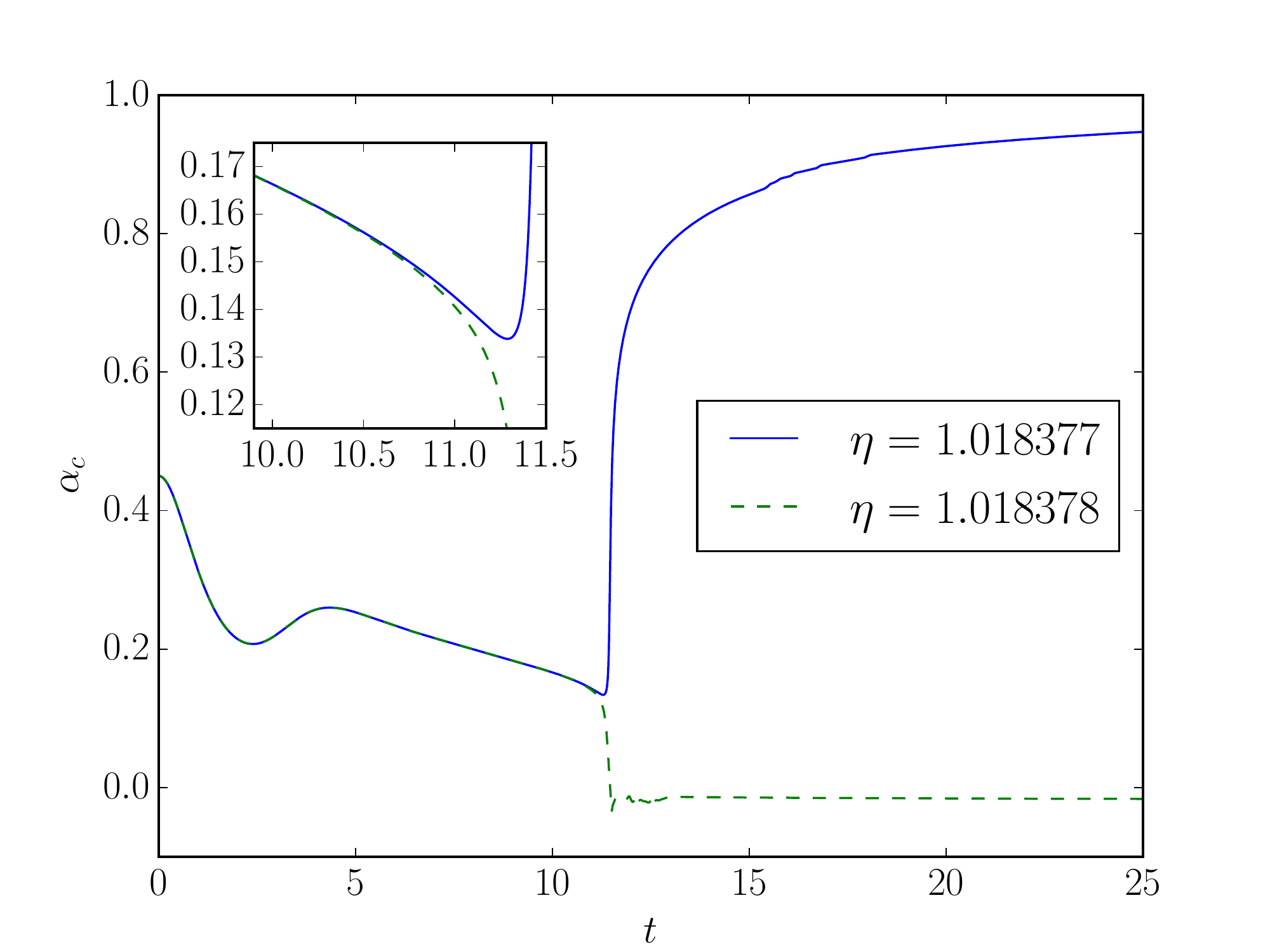}
\end{center}
% Produced with command lapse() in EC_Plots.py in /mnt/research/tbaumgar/Work/EvansColeman/SphericalSymmetry/CENTERED
\caption{The lapse function $\alpha$, interpolated to the center at $r = 0$, as a function of time $t$, for spherically symmetric ($\epsilon = 0$) and centered ($R_c = 0$) initial data.  We include results for two values of the parameter $\eta$ that bracket the critical value $\eta_c$.}
\label{Fig:lapse_centered}
\end{figure}

We start our discussion with spherically symmetric ($\epsilon = 0$) and centered ($R_c = 0$) data.   As explained above, the initial data (\ref{rho_init}) then reduce to those of Evans and Coleman \cite{EvaC94}, but we evolve these data with a completely independent numerical code, and different coordinate conditions.  For all simulations shown in this Section we used a "high-resolution" grid (see Section \ref{subsec:basics} for details.)

We perform numerical simulations for numerous different values of $\eta$ and, as expected, find that the simulations result in black-hole formation for sufficiently large values of this parameter.  In Fig.~\ref{Fig:lapse_centered}, for example, we show examples for two values of $\eta$ that, at least for the grid set-up used in these simulations, bracket the critical value $\eta_c$.  The two values of $\eta$ are so similar that, at early times, the two resulting curves are indistinguishable.  At late times, however, the lapse evolves very differently.  In the subcritical case ($\eta = 1.018377$) the lapse returns to approximately unity as the matter is dispersing to infinity, leaving behind flat space, while in the supercritical case ($\eta = 1.018378$) the lapse drops to approximately zero, which is an indication of black-hole formation.\footnote{We note that in Fig.~\ref{Fig:lapse_centered} we plot values that are interpolated to the origin at $r=0$.  This fourth-order interpolation using grid points on both sides of the origin leads to spuriously negative values after a black hole has formed, even though the lapse remains positive on all grid points.}  This bracketing of the critical parameter $\eta_c$ is in excellent agreement with the findings of \cite{EvaC94}, who reported $\eta_c \approx 1.0188$.

Unless the forming black hole is too small in comparison to our grid resolution, we can also detect the newly forming apparent horizon with our apparent horizon finder, and compute an approximate black-hole mass from the proper area of this apparent horizon.   For $\eta = 1.018378$, for example, we found that an apparent horizon forms at a coordinate radius of about 2.94 $\times 10^{-3}$, meaning that the interior of the black hole is covered by only 11 grid points (on our "high-resolution" grid.)  Accordingly, our error in the masses of small black holes is rather large.  For larger black holes, the apparent horizon grows a little after it first forms, as more matter falls into the black hole, but then settles down to an approximately constant value, which we take as the black hole mass $M_{\rm BH}$.

\begin{figure}
\begin{center}
\includegraphics[width=3.5in]{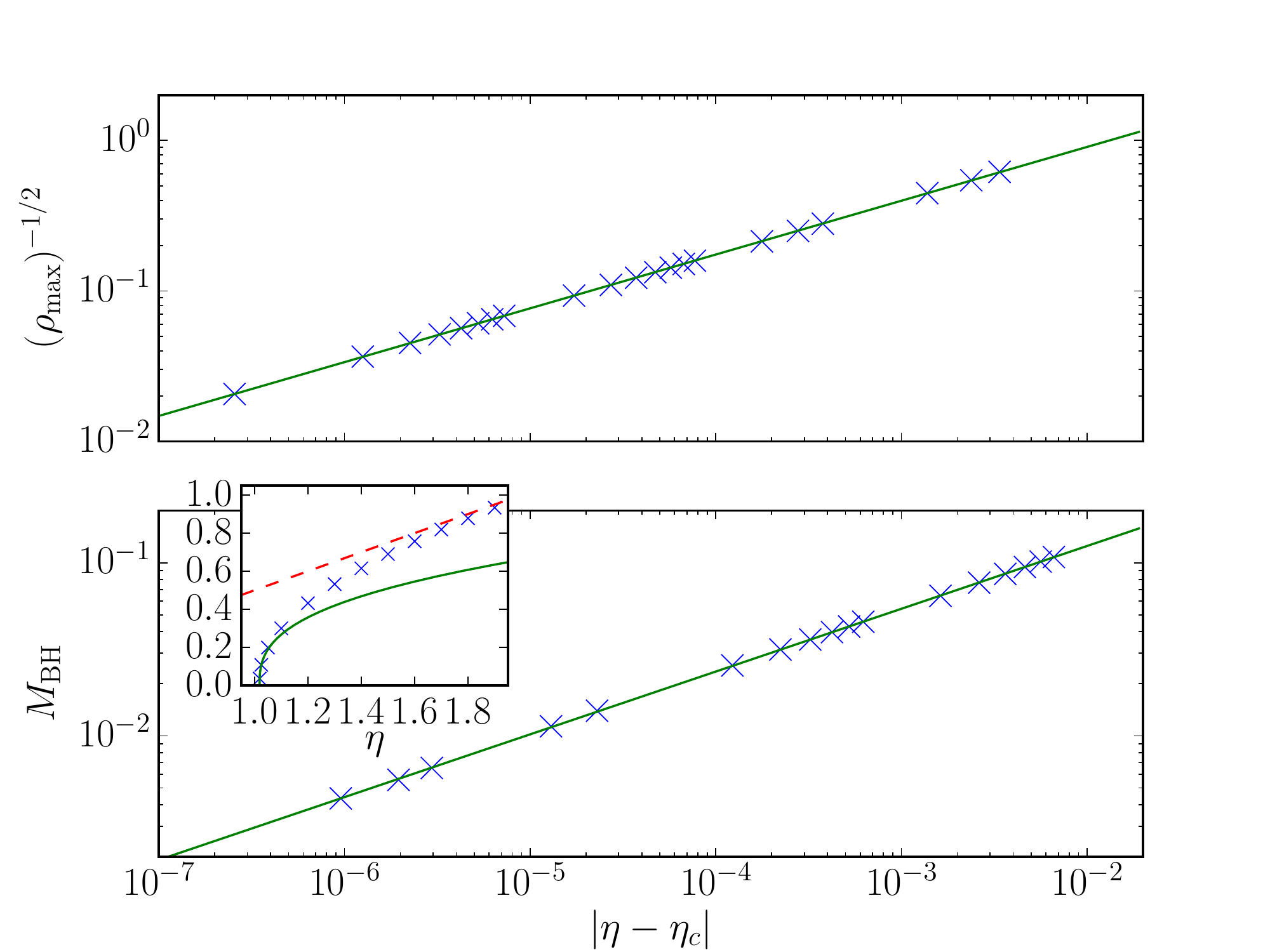}
\end{center}
% produced with command scaling_spherical_centered() in fits.py in /mnt/research/tbaumgar/Work/EvansColeman/FITS
\caption{Critical scaling of the maximum density $\rho^{\rm max}$ for subcritical evolutions (top panel), and the black hole mass $M_{\rm BH}$ for supercritical evolutions (bottom panel), for spherically symmetric ($\epsilon = 0$) and centered ($R_c = 0$) data.  The crosses denote our numerical results, while the solid lines represent the fits (\ref{mass_scaling}) and (\ref{rho_scaling}).  The dashed line in the inset shows the gravitational mass (\ref{ADM_mass}).}
\label{Fig:scaling_centered}
\end{figure}

In the bottom panel of Fig.~\ref{Fig:scaling_centered} we graph these black hole masses $M_{\rm BH}$ for supercritical evolutions ($\eta > \eta_c$) as a function of $\eta - \eta_c$.  Here we have determined the critical parameter $\eta_c = 1.0183770$ by fitting our numerical values to the scaling law 
\begin{equation} \label{mass_scaling} 
M_{\rm BH} = C_{\rm super} ( \eta - \eta_c)^\gamma,
\end{equation}
which is included in the figure as a solid line.  This fit also suggests $\gamma = 0.363$, in good agreement with the value of $\gamma = 0.36$ reported by \cite{EvaC94}.   We note, however, that our black hole masses are smaller than those reported in \cite{EvaC94} by about a factor of three.  Our code, including the apparent horizon finder, has passed many tests -- for example, we verified that the black-hole mass approaches that of the spacetime's gravitational mass (\ref{ADM_mass}) as $\eta$ is increased; see the inset in Fig.~\ref{Fig:scaling_centered}.   We are not aware of any problems, but we will continue to search for possible causes of this inconsistency.  In either case, none of our analysis in what follows relies on the masses of the forming black holes.

As pointed out by Garfinkle and Duncan \cite{GarD98}, the critical scaling exponent $\gamma$ can also be determined from the maximum value of the spacetime curvature attained in subcritical evolutions.  Since this curvature is related to the density $\rho$ by Einstein's equations (\ref{einstein}), we measure the maximum value of the density $\rho$ and, on dimensional grounds, fit to the scaling law
\begin{equation} \label{rho_scaling} 
\rho^{-1/2}_{\rm max} = C_{\rm sub} ( \eta_c - \eta)^\gamma.
\end{equation}
Our numerical results, together with the fit, are shown in the top panel of Fig.~\ref{Fig:scaling_centered}.  Using the data included in the plot, this fit results in $\eta_c = 1.0183773$ and $\gamma = 0.357$.    

Various sources of error contribute to the uncertainties in our reported values.  One source of error is the truncation error in our finite-difference calculation.  To estimate this error we performed simulations for a fixed value of $\eta = 1.0189$ with different resolutions and conclude that, for our high-resolution grid, the error in the resulting black-hole mass is a fraction of one percent.  Closer to the critical point, however, the solution develops smaller spatial features, so that our truncation error is larger (see also the discussion above.)  Fitting to the scaling laws (\ref{mass_scaling}) and (\ref{rho_scaling}) introduces additional systematic errors, as they hold strictly only in the immediate neighborhood of the critical point (see the inset in Fig.~\ref{Fig:scaling_centered}.)  Including or excluding data very close to the critical point (where numerical error will be larger) or further away from the critical point (where the scaling laws start to break down) will change our estimate for the critical exponent by a few percent.  We therefore estimate the error in the critical exponents -- at least those derived from the maximum density -- to be about 2 or 3\%.  This means that our results are in good agreement with those reported by \cite{EvaC94} ($\gamma \approx 0.36$) as well as the analytical results of \cite{KoiHA95,Mai96}, who found $\gamma \approx 0.3558$.   We believe that our results for the critical parameter $\eta_c$ are more accurate; in Section \ref{sec:ss_offcenter} below we compare results for both a high-resolution and a low-resolution grid and find agreement to within less than 0.1\%.  

\begin{figure}
\begin{center}
\includegraphics[width=3.5in]{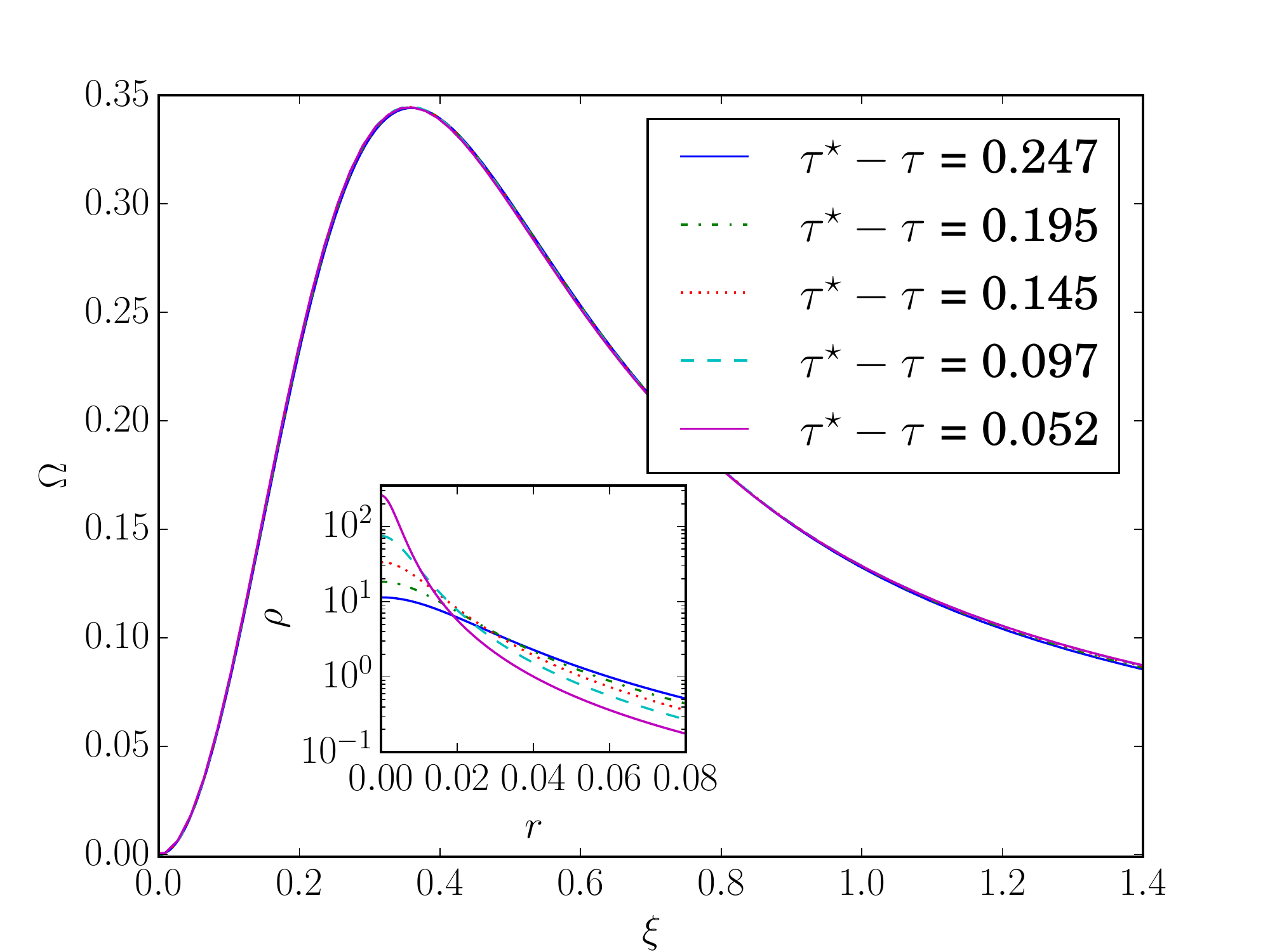}
\end{center}
\caption{The density function $\Omega(\xi)$ (see eq.~(\ref{Omega})) as identified from our numerical solution for $\eta = 1.0183772$ at coordinate times $t = 9.86$, 10.17, 10.49 10.80 and 11.11 (compare the inset in Fig.~\ref{Fig:lapse_centered}.)  The corresponding proper times at the center are $\tau = 2.377$, 2.429, 2.479, 2.527 and 2.572, and we have adopted $\tau^\star = 2.624$ in the construction of the self-similar variable $\xi$ (see eq.~(\ref{xi}).)  While the density $\rho$ itself changes significantly between these times (see the inset), the function $\Omega$ displays the expected self-similar behavior and depends on $\xi$ only (compare also Fig.~1 in \cite{EvaC94}.)}
\label{Fig:Omega_centered}
\end{figure}

We also examine the self-similarity of the critical solution.  In the strong-field region of the spacetime, close to criticality and prior to black-hole formation, as the spacetime ``tries to decide" whether or not to collapse to a black hole, the solution contracts to a focal point -- or rather ``focal event" -- in a self-similar fashion (see Fig.~1 in \cite{NeiC00b} for an illustration.)  In order to analyze this behavior we first define a variable
\begin{equation} \label{arealR}
R \equiv \psi^2 \left(\bar \gamma_{\theta\theta} \right)^{1/2} .
\end{equation}
In spherical symmetry (for which $\bar \gamma_{\phi\phi} = \sin^2 \theta \, \bar \gamma_{\theta\theta}$), $R$ becomes independent of $\theta$ and may be interpreted as the areal radius.  We then define a self-similar coordinate\footnote{We note that we consider here ``self-similarity of the first kind", rather than the second kind, which allows for a similarity exponent $n$ in the definition $\xi \equiv R/(\tau^* - \tau)^n$.  Evans and Coleman \cite{EvaC94} considered the latter, but also defined $\xi$ in terms of coordinate time rather than proper time at the center (see also the discussion in \cite{Gun03}.)}
\begin{equation} \label{xi}
\xi \equiv \frac{R}{\tau^* - \tau},
\end{equation}
where $\tau$ is the proper time measured by an observer at the origin, $r = 0$, and $\tau^*$ is the value of this proper time for the focal event (which, a priori, is not known.)   A self-similar solution can then be expressed as a function of $\xi$.  

As an example of a self-similar variable we follow \cite{EvaC94} and define a measure of the density,
\begin{equation} \label{Omega}
\Omega \equiv 4 \pi R^2 \rho.
\end{equation}
In Fig.~\ref{Fig:Omega_centered} we graph the function $\Omega$ versus $\xi$ for an evolution with $\eta = 1.0183772$, which is close to the critical parameter.  We construct $\Omega$ and $\xi$ from data at different times just prior to either black-hole formation or dispersal (see the inset in Fig.~\ref{Fig:lapse_centered}.)  Choosing $\tau^* = 2.624$ the resulting curves are so similar that they can hardly be distinguished in Fig.~\ref{Fig:Omega_centered} -- as expected for a self-similar solution.  

%%%%%%%%%%%%%%%%%%%%%%%%%%%%%%%%%%%%%%%%%%%%%%%%%%%%%%%%%%%%
\subsubsection{Off-center Data: $R_c = 3$}
\label{sec:ss_offcenter}
%%%%%%%%%%%%%%%%%%%%%%%%%%%%%%%%%%%%%%%%%%%%%%%%%%%%%%%%%%%%

\begin{figure}
\begin{center}
\includegraphics[width=3.5in]{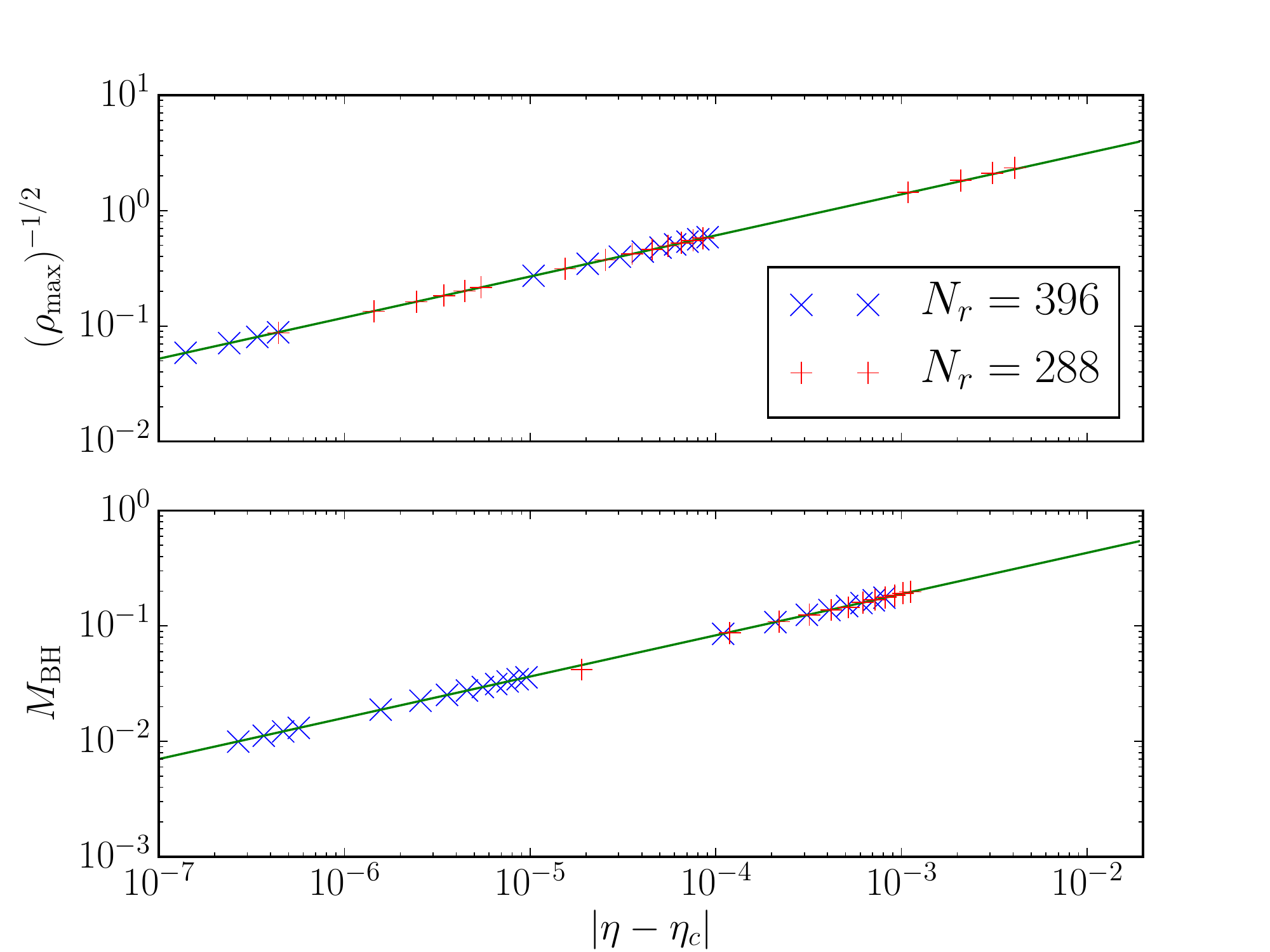}
\end{center}
% Produced with command scaling_spherical_offcenter() in fits.py in /mnt/research/tbaumgar/Work/EvansColeman/FITS
\caption{Same as Fig.~\ref{Fig:scaling_centered}, but for off-center ($R_c = 3$), spherically symmetric ($\epsilon = 0$) data.  Here we include numerical results obtained both on high and low-resolution grids.  The solid lines represent the fits (\ref{mass_scaling}) and (\ref{rho_scaling}), fitted to the high-resolution results.}
\label{Fig:scaling_offcentered}
\end{figure}

Before turning to axisymmetric simulations in Section \ref{sec:axi}, we discuss spherically symmetric ($\epsilon = 0$) off-center data with $R_c = 3$.  Qualitatively, the results are very similar to those for the centered data in Section \ref{sec:ss_centered}, except that now the fluid first has to propagate to the origin before it can collapse there.   The fluid therefore collapses at a later time than for the centered data.  As the fluid converges toward to the origin its density increases; accordingly, the critical parameter $\eta_c$ is smaller for off-center data than for centered data.  

In Fig.~\ref{Fig:scaling_offcentered} we show both subcritical and supercritical scaling for the off-centered data, similar to Fig.~\ref{Fig:scaling_centered} for centered data.  Here, however, we include results obtained both on a high-resolution and a low-resolution grid (see Section \ref{subsec:basics}.)    The solid line is a fit based on the high-resolution data; for the subcritical data we found the best fit for $\gamma = 0.356$, while for the supercritical data we found $\gamma = 0.357$ -- both in excellent agreement with the expected values.  For the high-resolution data we found $\eta_c \approx 0.1240904$, while for the low-resolution data $\eta_c \approx 0.124085$.

\begin{figure}
\begin{center}
\includegraphics[width=3.5in]{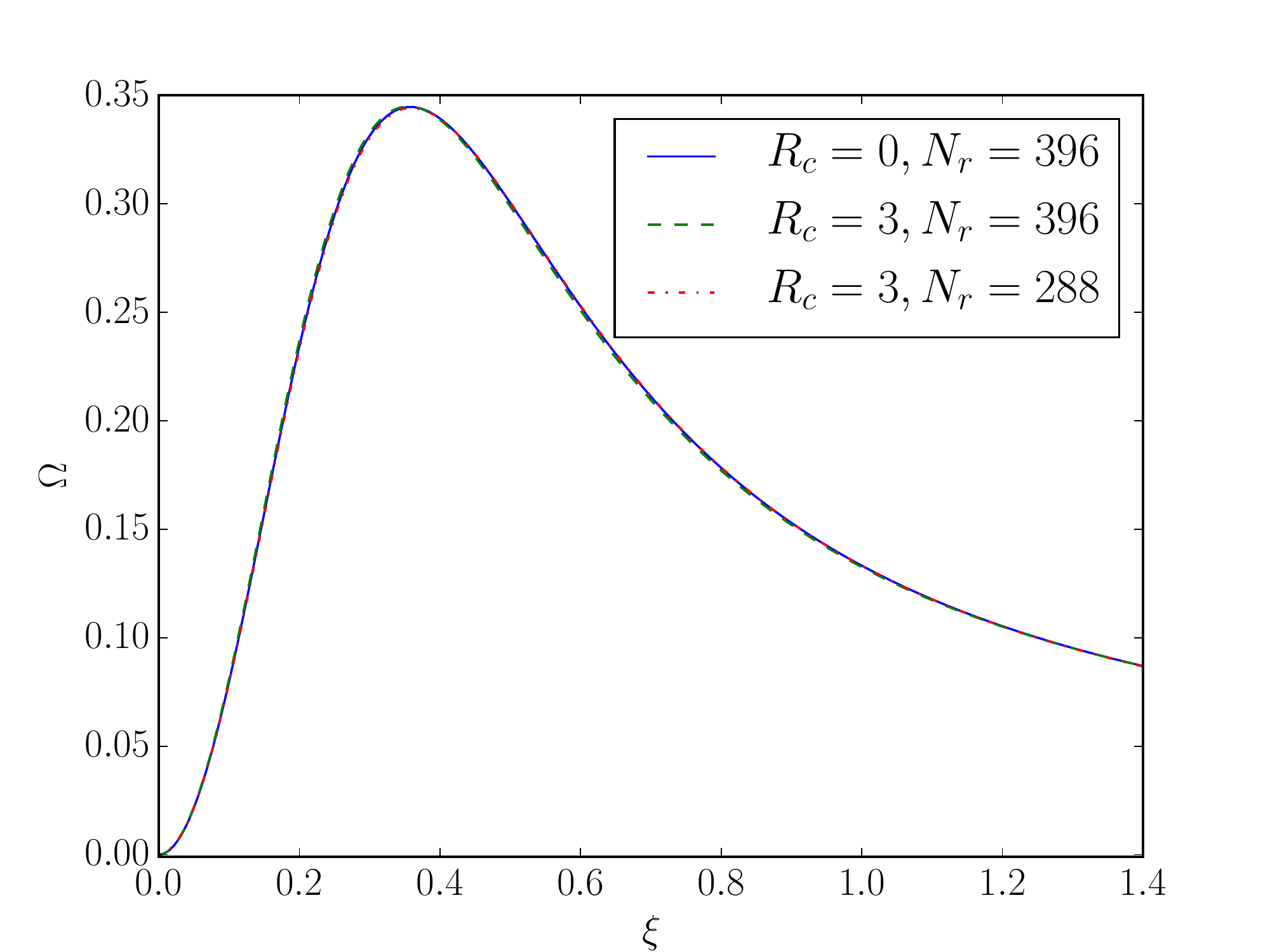}
\end{center}
% Produced with command self_sim_rho() in CompCritSol.py in /mnt/research/tbaumgar/Work/EvansColeman
\caption{Comparison of the self-similar variable $\Omega(\xi)$ as found in the evolution of centered initial data ($R_c = 0$, compare Fig.~\ref{Fig:Omega_centered}), with that found in the evolution of off-centered data ($R_c = 3$).  The two solutions agree very well, demonstrating the universality of the critical solution.  For the off-centered data we show results obtained on both high-resolution ($N_r = 396$) and low-resolution ($N_r = 288$) grids.}
\label{Fig:Omega_comp}
\end{figure} 

Evolutions closer to the critical parameter develop structures on a smaller spatial scale (in particular smaller black holes for supercritical evolutions) and hence require higher resolution; it is therefore not surprising that simulations with a higher radial resolution remain reliable closer to the critical parameter.  While we cannot perform simulations quite as close to the critical parameter on the low-resolution grid as with the high-resolution grid, it is evident from Fig.~\ref{Fig:scaling_offcentered} that the low-resolution grid is still perfectly adequate for the observation of critical scaling.   

Similar to our analysis in Section \ref{sec:ss_centered} we also identify the critical solution in these evolutions; adopting a proper time $\tau^* = 6.45$ for the focal event our results are very similar to those shown for the centered data in Fig.~\ref{Fig:Omega_centered}.  In Fig.~\ref{Fig:Omega_comp} we show a comparison between the critical solutions found in the centered and off-centered simulations.  The two solutions can hardly be distinguished, which demonstrates the universality of the critical solution.  In the same plot we also include the critical solution as observed on a low-resolution grid.  The excellent agreement demonstrates that the low-resolution  grid is sufficient to identify the critical solution, and we therefore will use a low-resolution grid only in the axisymmetric simulations of Section \ref{sec:axi}.

%%%%%%%%%%%%%%%%%%%%%%%%%%%%%%%%%%%%%%%%%%%%%%%%%%%%%%%%%%%%
\subsection{Axiymmetry: $\epsilon = 0.1$ and $\epsilon = 0.5$}
\label{sec:axi}
%%%%%%%%%%%%%%%%%%%%%%%%%%%%%%%%%%%%%%%%%%%%%%%%%%%%%%%%%%%%

\begin{figure}
\begin{center}
\includegraphics[width=3.5in]{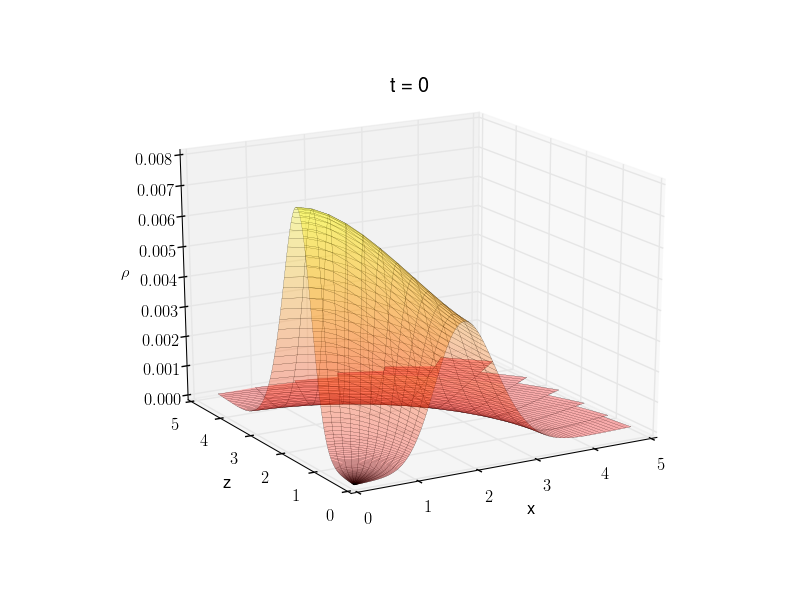}
\includegraphics[width=3.5in]{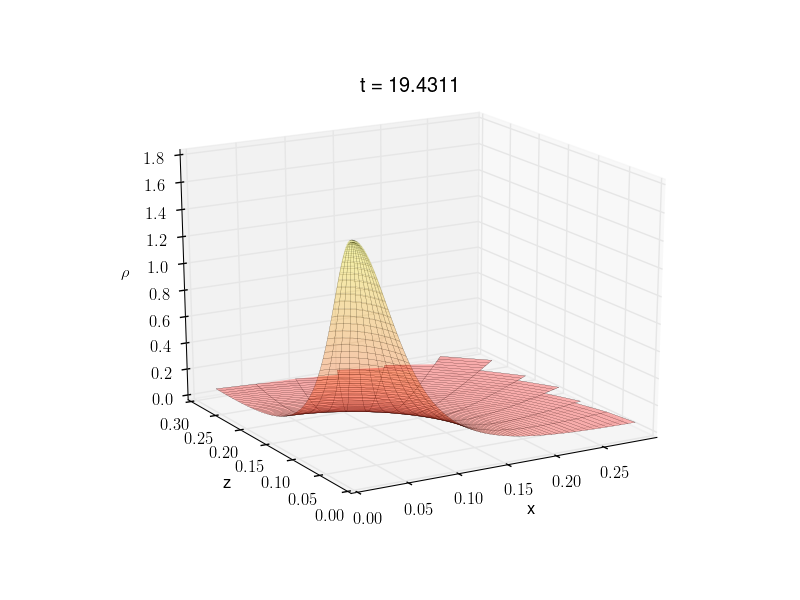}
\includegraphics[width=3.5in]{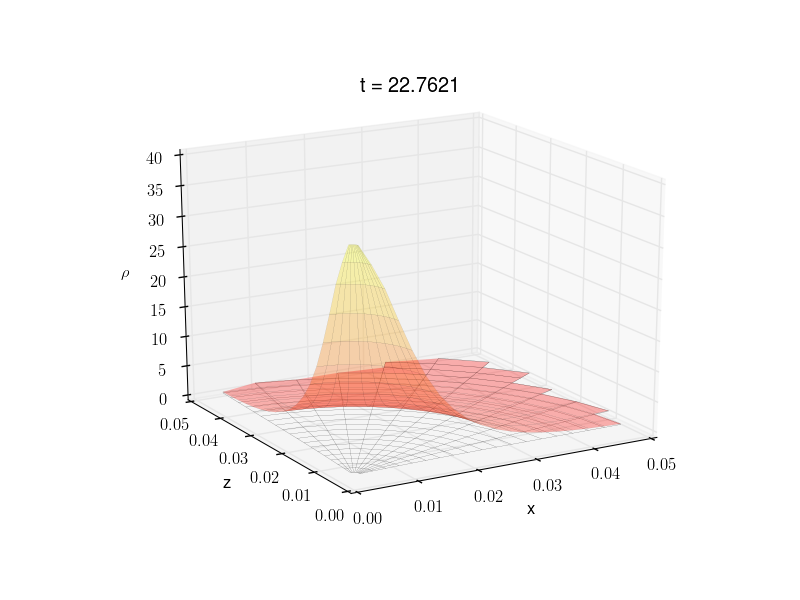}
\end{center}
% Produced with command compare() in data_reader.py in /mnt/research/tbaumgar/Work/EvansColeman/AxiSymmetry/OFF_CENTER/EPS_05 
\caption{Profiles of the density $\rho$ at different instances of coordinate time $t$ for data with $\epsilon = 0.5$ and $R_c = 3$.  The black wireframe shows the density for a subcritical evolution with $\eta = 0.12442$, while the colored surface shows results for a supercritical evolution with $\eta = 0.12443$.  In the first two panels the two surfaces cannot be distinguished, but in the last panel the density keeps increasing for the supercritical evolution, while for the subcritical evolution the density has dropped to very small values.  Note the vastly different scales in the different panels.}
\label{Fig:rho_axi}
\end{figure} 

\begin{figure}
\begin{center}
\includegraphics[width=3.5in]{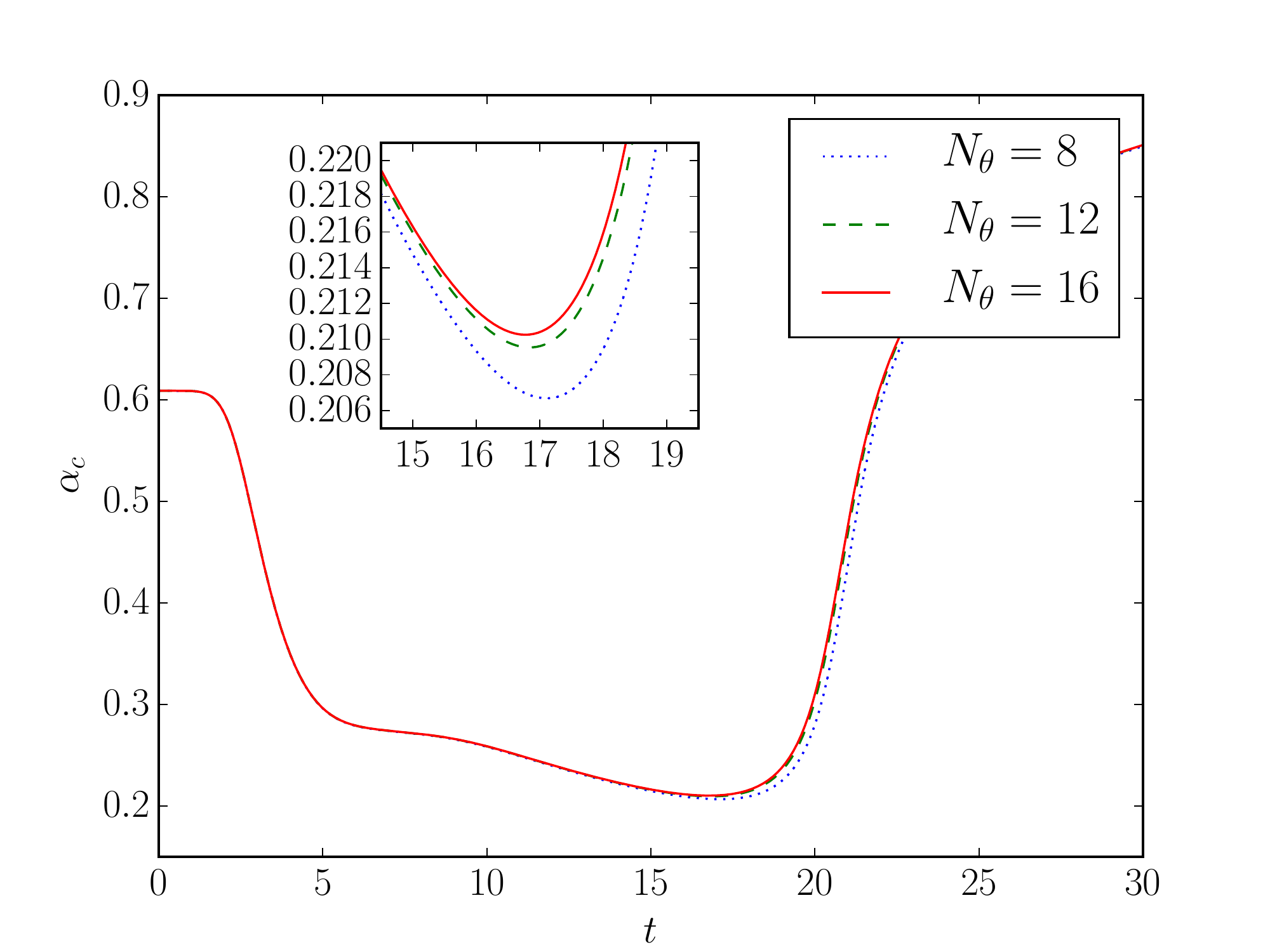}
\end{center}
% Produced with command converge() in convergence.py in /mnt/research/tbaumgar/Work/EvansColeman/CONVERGENCE
\caption{The central value of the lapse $\alpha_c$ for a subcritical ($\eta = 0.1240$) axisymmetric ($\epsilon = 0.5$) evolution for three different angular resolutions, $N_\theta = 8$, 12 and 16.  Even though these are all quite coarse resolutions, we see that the difference between the two higher resolutions is already relatively small.} 
\label{Fig:ang_convergence}
\end{figure}

We now turn to the aspherical collapse of radiation fluids, and consider evolutions with $\epsilon = 0.1$ and 0.5.  All results presented in this section are for off-centered data with $R_c = 3$, evolved with a low-resolution radial grid with $N_r = 288$.

As an example of our evolutions we show profiles of the density $\rho$ at three different coordinate times in Fig.~\ref{Fig:rho_axi}.  We show results for a subcritical evolution with $\eta = 0.12442$ (represented as a wireframe), as well as a supercritical evolution with $\eta = 0.12443$ (represented as a colored surface), both for $\epsilon = 0.5$.   At the earlier two times the two profiles cannot be distinguished, but at the latest time the density for the supercritical evolution continues to increase, while the density for the subcritical evolution has dropped down to much smaller values.   These simulations were carried out with a very modest angular resolution of $N_\theta = 12$ (since we impose equatorial symmetry, these grid points cover only one hemisphere.)    In order to evaluate whether such a coarse resolution is adequate, we show in Fig.~\ref{Fig:ang_convergence} results for a subcritical evolution with $\eta = 0.1240$ and $\epsilon = 0.5$ for three different resolutions, $N_\theta = 8$, 12 and 16 grid points.  Even though these are still very small resolutions, we see that the difference in the two higher resolutions is relatively small, in part because the deviations from sphericity remain relatively small throughout the evolution.   As a compromise between accuracy and computing time we therefore chose $N_\theta = 12$ for all simulations presented in this Section. 

\begin{figure}
\begin{center}
\includegraphics[width=3.5in]{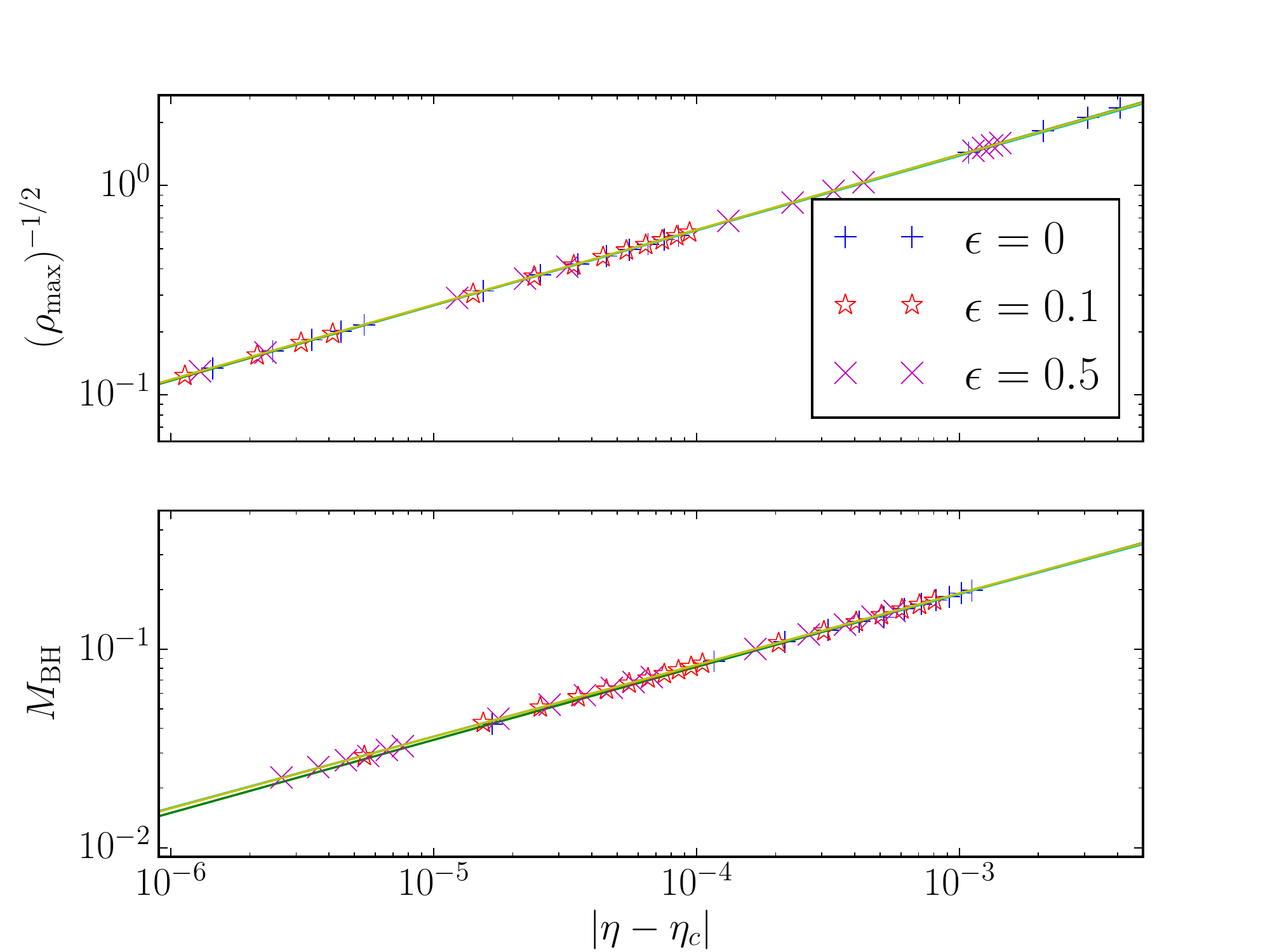}
\end{center}
% Produced with command scaling_axi() in fits.py in /mnt/research/tbaumgar/Work/EvansColeman/FITS
\caption{Same as Fig.~\ref{Fig:scaling_offcentered}, but for both spherical and axisymmetric data.  The solid lines represent the fits (\ref{mass_scaling}) and (\ref{rho_scaling}).  The spherical data for $\epsilon = 0$ are the same as the low-resolution data in Fig.~\ref{Fig:scaling_offcentered}.}
\label{Fig:scaling_axi}
\end{figure}

In Fig.~\ref{Fig:scaling_axi} we show the critical scaling of the maximum density (for subcritical evolutions) and black-hole mass (for supercritical evolutions) in our simulations.  The spherical data for $\epsilon = 0$ are identical to the low-resolution results shown in Fig.~\ref{Fig:scaling_offcentered}.  The aspherical data for $\epsilon = 0.1$ and 0.5 follow remarkably similar scaling laws for both the maximum density and the black-hole masses; fits to these different data sets can hardly be distinguished in Fig.~\ref{Fig:scaling_axi}.   To within the accuracy of our simulations, the critical exponent $\gamma$ appears to be the same for our choices of $\epsilon$ -- it is possible, of course, that larger deformations from sphericity would lead to a change in $\gamma$ (compare \cite{ChoHLP03b}.)   The critical parameter $\eta_c$, on the other hand increases slightly with $\epsilon$; for $\epsilon = 0.1$ we found $\eta_c \approx 0.124094$, and for $\epsilon = 0.5$ about $\epsilon \approx  0.124432$, compared to about $\eta_c \approx 0.124085$ for the the low-resolution spherical data with $\epsilon = 0$. 

\begin{figure}
\begin{center}
\includegraphics[width=3.5in]{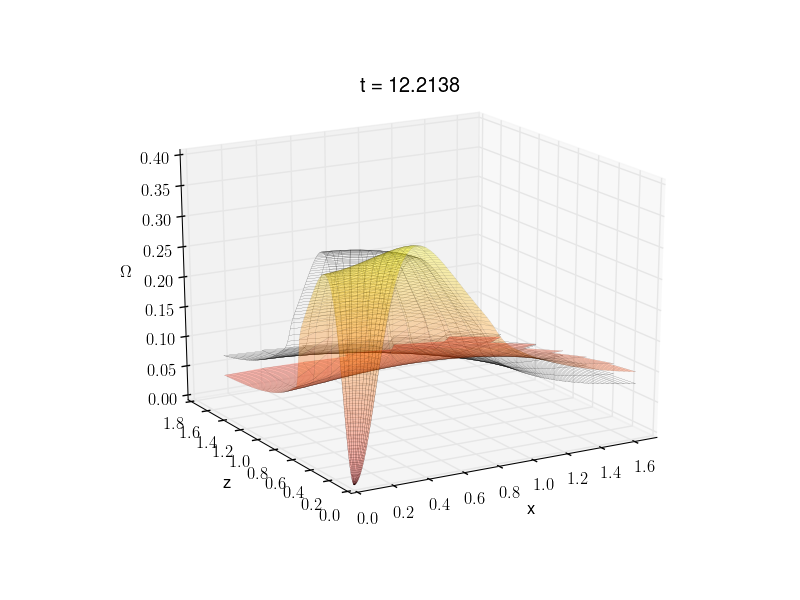}
\includegraphics[width=3.5in]{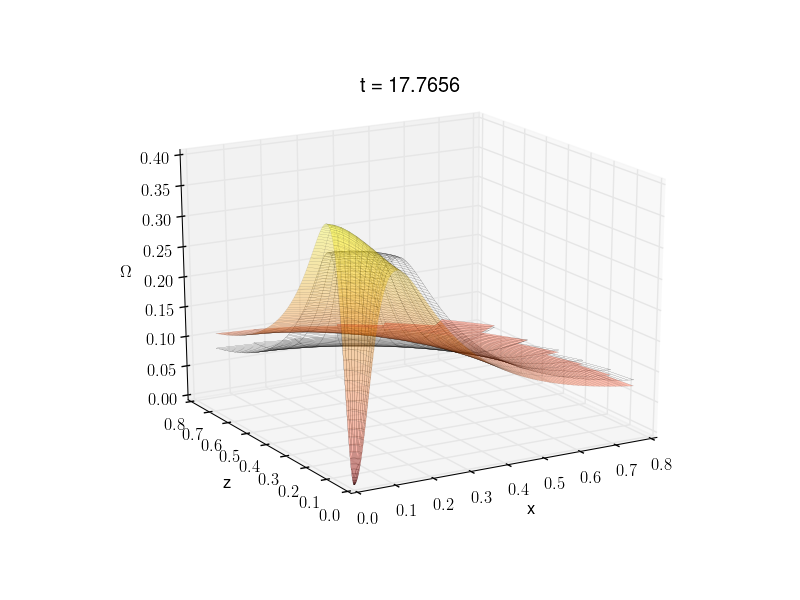}
\includegraphics[width=3.5in]{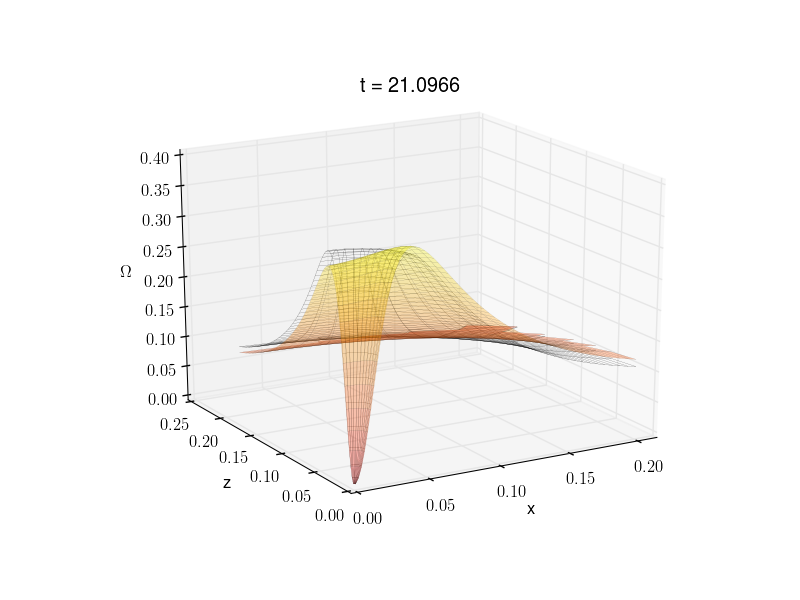}
\end{center}
% Produced with command compare_Omega() in data_reader.py in /mnt/research/tbaumgar/Work/EvansColeman/CRIT_SOLS/OFF_CENTERED/MED_RES
\caption{Profiles of the function $\Omega$ at three different instances of coordinate time $t$ for near-critical evolutions.  The colored surface shows results for $\epsilon = 0.5$ and $\eta = 0.124432$, while the wireframe shows results for spherical data with $\epsilon = 0$ and $\eta = 0.124085$ (see Fig.~\ref{Fig:Omega_comp}.)  The aspherical collapse leads to damped oscillations around the critical solution; we show these oscillations approximately at times when their difference from sphericity is greatest.}
\label{Fig:Omega_profiles}
\end{figure} 

We now turn to the approach of aspherical collapse to the critical solution.  In Fig.~\ref{Fig:Omega_profiles} we show profiles of the function $\Omega$, defined in eq.~(\ref{Omega}), for a spherical and aspherical evolution close to criticality.  In these profiles, the colored surfaces show profiles of $\Omega$ for an aspherical collapse with $\epsilon = 0.5$, while the wireframes show profiles for a spherical collapse.   The radial contours for the spherical collapse are similar to those shown in Figs.~\ref{Fig:Omega_centered} and \ref{Fig:Omega_comp}, except that $\Omega$ is shown as a function of the self-similar variable $\xi$ there, while we graph $\Omega$ as a function of $x = r \cos \theta$ and $y = \sin \theta$ here.  In the aspherical evolution the density function $\Omega$ appears to ``slosh" back and forth between the poles and the equator in  damped oscillations around the spherical critical solution.  In Fig.~\ref{Fig:Omega_profiles} we show these oscillations at times that more or less correspond to the largest deviations from sphericity.  

\begin{figure}
\begin{center}
\includegraphics[width=3.5in]{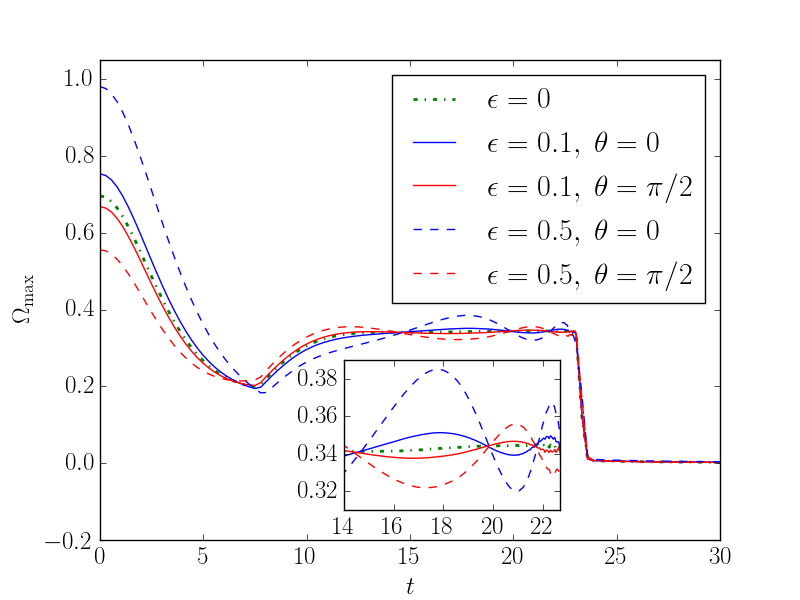}
\end{center}
% Produced with command Omega() in find_max.py in /mnt/research/tbaumgar/Work/EvansColeman/AxiSymmetry/OFF_CENTER/ 
\caption{Maxima of the function $\Omega$ along the axis ($\theta = 0$) and in the equatorial plane ($\theta = \pi/2$), as a function of time, for subcritical evolutions close to the critical parameter $\eta_c$.  The spherical self-similar contraction shown in Fig.~\ref{Fig:Omega_comp} occurs while the $\Omega_{\rm max}$ is approximately constant.  Here we include results for $\epsilon = 0.1$ and $0.5$; it can be seen that these maxima oscillate around the respective spherical values.}
\label{Fig:Omega_max}
\end{figure} 

The oscillations can be seen more clearly in Fig.~\ref{Fig:Omega_max}, where we show the maxima of $\Omega$, both along the axis and in the equatorial plane, as a function of time.  The spherical data form a plateau between times of approximately $t =12$ and $t=23$; it is during this time that the solution contracts in a self-similar fashion as shown in Fig.~\ref{Fig:Omega_comp}.  For the aspherical evolutions we observe oscillations around these spherical results; these oscillations appear to be damped, and to have a decreasing period, as one might expect from an oscillation around a self-similarly contracting solution.  

\begin{figure}
\begin{center}
\includegraphics[width=3.5in]{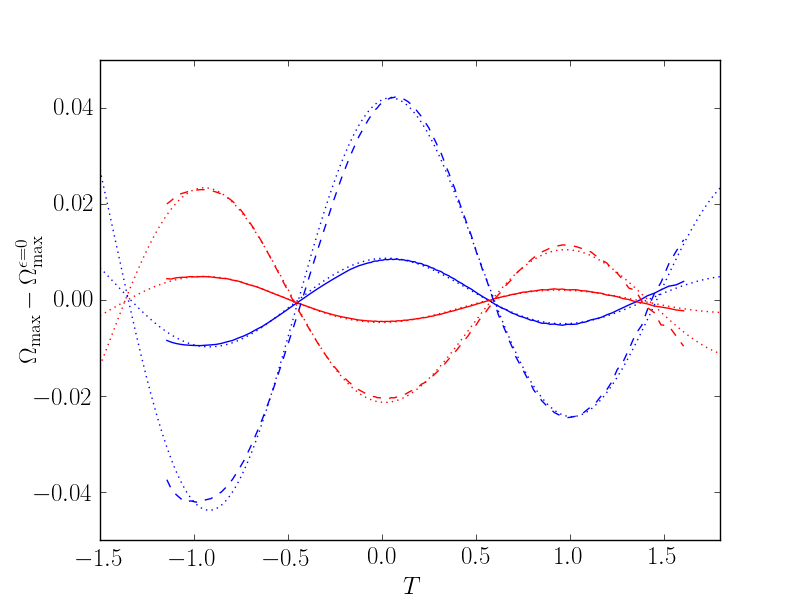}
\end{center}
% Produced with command Omega() in find_max.py in /mnt/research/tbaumgar/Work/EvansColeman/AxiSymmetry/OFF_CENTER/ 
\caption{Same as Fig.~\ref{Fig:Omega_max}, with the same symbols used for the respective lines, except that here we show the differences between the aspherical and the spherical data.  We also show these data as a function of the time coordinate $T$ (see eq.~(\ref{T})), in terms of which the focal event occurs at positive infinity.  The dotted lines are fits based on (\ref{mg}).}
\label{Fig:Omega_diff}
\end{figure}

In order to analyze these oscillations more quantitatively, we plot in Fig.~\ref{Fig:Omega_diff} the differences in the maximum values of $\Omega$ between aspherical and spherical data.  We also display these differences as a function of a time coordinate 
\begin{equation} \label{T}
T = -\ln( \tau^* - \tau ), 
\end{equation}
appropriate for self-similar collapse.  In terms of this coordinate, the focal event at $\tau = \tau^*$, which we take to be $\tau^* = 6.45$ as found in Section \ref{sec:ss_offcenter}, occurs at a time of positive infinity.  Gundlach \cite{Gun98b,Gun02} found that perturbations of a critical solution perform damped oscillations that, up to another periodic function, can then be written in the form
\begin{equation} \label{mg}
u(T) = A e^{- \kappa T} \cos(\omega T + \phi). 
\end{equation}
Here $\kappa$ is a damping coefficient, $\omega$ the frequency of the oscillations, and $\phi$ a phase shift.   

In order to test this, we include in Fig.~\ref{Fig:Omega_diff} fits based on (\ref{mg}) as the dotted lines.\footnote{Strictly speaking, we include in these fits an additional parameter $u_0$ that allows for a non-zero off-set in eq.~(\ref{mg}), so that $u(T) - u_0$ equals the right-hand side of (\ref{mg}).  We always found this parameter to be small.}  For $\epsilon = 0.1$ we obtain excellent fits with values of $\omega$  within one percent of $\omega = 3.33$ for both the axis and equatorial data, while the values for $\kappa$ are within 10\% of $\kappa = 0.35$.   The fits for $\epsilon = 0.5$ are not quite as clean as those for $\epsilon = 0.1$ -- possibly these deviations are caused by nonlinear effects.   The values of $\omega$ based on fits for $\epsilon = 0.5$ are still very similar to those found for $\epsilon = 0.1$, but the value of $\kappa$ found from a fit for the axis data is somewhat smaller ($\kappa \approx 0.27$.)  We caution, however, that the exact value of these fits again depends on how early or late data points we include.  Earlier data points are affected by the solution not having entered the self-similar contraction yet, while later data points show more numerical noise as the solution contracts to very small spatial scales and we lose sufficient resolution.  

Quite reassuringly, the above values of $\omega$ and $\kappa$ are close to those found by Gundlach \cite{Gun02} for aspherical perturbations of the critical collapse of perfect fluids.  For polar $\ell \geq 2$ perturbations (see his Section IV.F), and for a radiation fluid, 
his results can be read off from his Fig.~11, with $\omega \approx 3.6$ and $\kappa \approx 0.37$.  Both values are within about 10\% of the values that we identified from our fully nonlinear simulations.

%%%%%%%%%%%%%%%%%%%%%%%%%%%%%%%%%%%%%%%%%%%%%%%%%%%%%%%%%%%%
\section{Discussion}
\label{sec:discussion}
%%%%%%%%%%%%%%%%%%%%%%%%%%%%%%%%%%%%%%%%%%%%%%%%%%%%%%%%%%%%

We perform numerical simulations of radiation fluids close to the onset of black-hole formation and study critical phenomena for spherically symmetric as well as aspherical initial data.   We find critical scaling for both subcritical and supercritical evolutions, and for both spherical and aspherical data.  We also identify the critical solution in spherical evolutions, find evidence for its universality, and demonstrate how, in aspherical evolutions, near-critical data perform a damped oscillation around the spherical critical solution -- at least in the cases we consider.

Our results are consistent with those of Gundlach \cite{Gun02}, who used perturbation techniques to show that all aspherical perturbations of the critical solution decay in damped oscillations (see also \cite{Gun98b}, as well as \cite{MarG99} for similar calculations for scalar fields.)  Matching our numerical results to damped oscillations of the form (\ref{mg}) we find damping coefficients and frequencies that are within about 10\% of those reported in \cite{Gun02}.   However, our results also do not rule out the existence of a growing nonspherical mode as reported by Choptuik {\it et.al.} \cite{ChoHLP03b}, since those modes may only appear for larger deviations from sphericity, closer to the critical point, or perhaps they may appear only for some matter models, but not for others.  It is also possible that the grid resolutions adopted in this paper are not sufficient to detect this mode.  We plan to further pursue this issue in future studies.

In addition to studying critical phenomena in the aspherical collapse of radiation fluids, this paper serves as a demonstration that an unconstrained evolution code, using "moving-puncture" coordinates, is suitable for the study of critical collapse, at least for some matter models (see also \cite{HilBWDBMM13,HilWB15,AkbC15} for recent discussions of this issue.)   
 
\acknowledgements

It is a pleasure to thank Steve Liebling and Frans Pretorius for helpful conversations, and Chuck Evans both for conversations and for sharing numerical data from \cite{EvaC94}.   We would also like to thank Carsten Gundlach for pointing us to reference \cite{Gun02}, and for several very useful comments.  TWB would like to thank the Max-Planck-Institut f\"ur Astrophysik in Garching (Germany) for its hospitality.  This work was supported in part by NSF grant PHY-1402780 to Bowdoin College, and by the Max-Planck-Institut f\"ur Astrophysik. 

\begin{appendix}
%%%%%%%%%%%%%%%%%%%%%%%%%%%%%%%%%%%%%%%%%%%%%%%%%%%%%%%%%%%%
\section{Logarithmic Finite-Difference Stencils}
\label{app:log}
%%%%%%%%%%%%%%%%%%%%%%%%%%%%%%%%%%%%%%%%%%%%%%%%%%%%%%%%%%%%

In this appendix we describe the structure of our logarithmic radial grid together with some of the finite-difference stencils that we use.

\begin{figure}
\begin{center}
\includegraphics[width=3.2in]{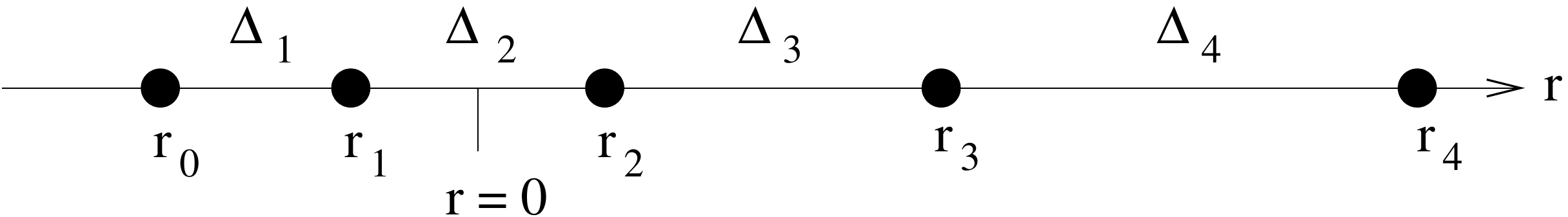}
\end{center}
\caption{Schematic drawing of the setup of our logarithmic grid in the radial direction.}
\label{Fig:grid}
\end{figure}

We denote gridpoints with $r_i$ and define
\begin{equation}
\Delta_i \equiv r_i - r_{i-1}
\end{equation}
(see the schematic drawing in Fig.~\ref{Fig:grid}.)  Since we use finite-difference stencils that, for each gridpoint, use up to two nearest neighbors on both sides, we define the first two ghost-points, $r_0$ and $r_1$, with negative values of the radius (see also Fig.~1 in \cite{BauMCM13}.)  We further choose the location of $r_1$ and $r_2$ so that the origin, at $r=0$, is half-way between these two points, i.e.
\begin{equation}
r_2 = - r_1 = \Delta_2/2.
\end{equation}
For a uniform grid the grid spacing $\Delta r_i$ is constant across the entire grid.  Here we instead allow for a logarithmic grid and choose
\begin{equation}
\Delta_{i+1} = c \Delta_i
\end{equation}
with $c \geq 1$.  

\begin{table*}
\begin{tabular}{c|c|c}
$I$	        &   $A_I$   & $B_I$ \\ 
	        \hline 
	        \hline
$+2$		&   $\displaystyle - \frac{1}{c^2 (1+c)(1+c^2)(1+c+c^2)}$ & $ \displaystyle \frac{2(1 - 2 c^2)}{c^3 (1+c)^2 (1+c^2) (1+c+c^2)} $\\
\hline 
$+1$	  	&  $\displaystyle \frac{1+c}{c^2(1+c+c^2)}$	& $\displaystyle \frac{2(2c - 1)(1+c)}{c^3(1+c+c^2)}$ \\
\hline
$0$ 		& $\displaystyle \frac{2(c - 1)}{c}$ 			& $\displaystyle \frac{2(1-c-5c^2-c^3 + c^4)}{c^2(1+c)^2}$ \\
\hline
$-1$		& $\displaystyle - \frac{c^2(1+c)}{(1+c+c^2)}$	& $\displaystyle \frac{2 c (2 - c)(1+c)}{(1+c+c^2)}$ \\
\hline
$-2$		& $\displaystyle \frac{c^6}{(1+c)(1+c^2)(1+c+c^2)}$ & $ \displaystyle \frac{2 c^5 (c^2 - 2)}{(1+c)^2 (1+c^2) (1+c+c^2)} $\\
\end{tabular}
\caption{Finite-difference coefficients $A_I$ and $B_I$ for the first (middle column) and second (right column) derivative on a logarithmic grid, using centered, five-point stencils (see eqs.~(\ref{first_deriv}) and (\ref{second_deriv}).)}
\label{table:stencil}
\end{table*}

We would now like to represent derivatives of a function $f$, say, in terms of finite-differencing stencils involving the function values at a  gridpoint $r_i$ as well as its two nearest neighbors on both sides.  To this end, we express the function values at those gridpoints in terms of a Taylor expansion up to fourth order about the central point, and write all distances in terms of $\Delta r_i$.   For $f_{i+2} = f(r_{i+2})$, for example, we have
\begin{eqnarray} \label{taylor}
f_{i+2} & = & f_i^{(0)} + \delta_{2} f_i^{(1)} 
+ \frac{1}{2!} \delta_{2}^2 f_i^{(2)}
+ \frac{1}{3!} \delta_{2}^3 f_i^{(3)}
+ \frac{1}{4!} \delta_{2}^4 f_i^{(4)} \nonumber \\
& & + \mathcal{O}(\delta_{2}^5 f_i^{(5)}),
\end{eqnarray}
where
\begin{equation}
\delta_{2} \equiv r_{i+2} - r_{i} =  c (1+c)  \Delta_i,
\end{equation}
and where $f_i^{(n)}$ denotes the $n$-th derivative of the function $f$ evaluated at $r = r_i$.  We write similar expressions for $f_{i+1}$, $f_{i-2}$ and $f_{i-2}$, as well as $f_i = f_i^{(0)}$, and observe that we can combine the resulting five equations into a single matrix equation
\begin{equation} \label{matrix_equation}
\left(
\begin{array}{c}
f_{i+2}^{} \\
f_{i+1}^{} \\
f_{i} \\
f_{i-1} \\
f_{i-2} \\
\end{array}
\right) 
= 
M 
\left(
\begin{array}{c}
f_i^{(0)}  \\
f_i^{(1)}  \\
f_i^{(2)}  \\
f_i^{(3)}  \\
f_i^{(4)}  \\
\end{array}
\right),
\end{equation}
where the matrix $M$ is given by
\begin{widetext}
\begin{equation}
M = 
\left(
\begin{array}{ccccc}
1  & \displaystyle c (1+c) \Delta_i 
& \displaystyle \frac{(c (1+c))^2}{2} \Delta_i^2 
& \displaystyle \frac{(c (1+c))^3}{6} \Delta_i^3 
& \displaystyle \frac{(c (1+c))^4}{24} \Delta_i^4 \\[2mm]
1  & \displaystyle c \Delta_i 
& \displaystyle \frac{c^2}{2} \Delta_i^2 
& \displaystyle \frac{c^3}{6} \Delta_i^3 
& \displaystyle \frac{c^4}{24} \Delta_i^4 \\ 
1 & 0 & 0 & 0 & 0 \\
1  & \displaystyle - \Delta_i 
& \displaystyle \frac{1}{2} \Delta_i^2 
& \displaystyle - \frac{1}{6} \Delta_i^3 
& \displaystyle \frac{1}{24} \Delta_i^4 \\[2mm] 
1  & \displaystyle - \frac{(1+c)}{c} \Delta_i 
& \displaystyle \frac{(1+c)^2}{2c^2} \Delta_i^2 
& \displaystyle \frac{(1+c)^3}{6c^3} \Delta_i^3 
& \displaystyle \frac{(1+c)^4}{24c^4} \Delta_i^4 
\end{array}
\right).
\end{equation}
\end{widetext}
We now invert the matrix equation (\ref{matrix_equation}) to obtain
\begin{equation} \label{inverse}
\left(
\begin{array}{c}
f_i^{(0)} \\
f_i^{(1)} \\
f_i^{(2)} \\
f_i^{(3)} \\
f_i^{(4)}  \\
\end{array}
\right) 
= M^{-1} 
\left(
\begin{array}{c}
f_{i+2}^{} \\
f_{i+1}^{} \\
f_{i} \\
f_{i-1} \\
f_{i-2} \\
\end{array}
\right),
\end{equation}
from which we can read off finite-difference expressions for the first four derivatives of the function $f$ in terms of the function values at $r_i$ and its nearest neighbors.   In particular, we have 
\begin{equation} \label{first_deriv}
f_i^{(1)} = \frac{1}{\Delta_i} \sum_{I=-2}^{+2} A_I f_{i+I}
\end{equation}
and 
\begin{equation} \label{second_deriv}
f_i^{(2)} = \frac{1}{\Delta_i^2} \sum_{I=-2}^{+2} B_I f_{i+I}
\end{equation}
for the first two derivatives, where the coefficients $A_I$ and $B_I$ are listed in Table \ref{table:stencil} (see also \cite{Den14}). 

For a uniform grid, with $c=1$, the finite-difference stencils reduce to the more familiar expressions
\begin{equation} \label{first_uniform}
f_i^{(1)} = \frac{1}{12 \Delta_i}(f_{i-2} - 8 f_{i-1} + 8 f_{i+1} - f_{i+2})
\end{equation}
and
\begin{equation} \label{second_uniform}
f_i^{(2)} = \frac{1}{12 \Delta_i^2}(-f_{i-2} + 16 f_{i-1} - 30 f_i + 16 f_{i+1} - f_{i+2}).
\end{equation}
For uniform girds the leading-order error terms in the Taylor expansion (\ref{taylor}) cancel out exactly when they are combined to form the expressions (\ref{first_uniform}) and (\ref{second_uniform}), so that in both of these expressions the error scales with $\Delta_i^4$.  This cancellation does not occur for a logarithmic grid with $c \ne 1$, so that the leading-order error term is one order lower.

We use a similar approach to derive expressions for one-sided derivatives that are used in up-wind differencing of advective shift terms.

\end{appendix}
%merlin.mbs apsrev4-1.bst 2010-07-25 4.21a (PWD, AO, DPC) hacked
%Control: key (0)
%Control: author (8) initials jnrlst
%Control: editor formatted (1) identically to author
%Control: production of article title (-1) disabled
%Control: page (0) single
%Control: year (1) truncated
%Control: production of eprint (0) enabled
%

% \bibliography{references}

\end{document}